\documentstyle[12pt,titolo]{article}
\def\bfm#1{{\mbox{\boldmath $#1$}}}
\begin{document}
\bibliographystyle{unsrt}
\title{The y-scaling in the $\Delta$-region}
\author{R.Cenni and P.Saracco}
\abstract{A scaling variable for electron scattering off nuclei in the
$\Delta$-region is shown to exist. A scaling law for
a large variety of scattering
data with different kinematics is found. This property turns out to be
very sensitive to the reaction mechanism. We will show that this outcome
follows from drastic medium induced modifications on the
momentum and energy dependence of the $\Delta$ width.
\vskip1.truecm\noindent \rm Revised version }
\maketitle


\section{Introduction}

In the past years electron scattering off nuclei has been extensively
investigated in terms of $y$-scaling \cite{DaMcDoSi-90,CiPaSa-86,CiPaSa-91},
since the pioneering work of West\cite{We-75} and the first analysis of
the $(e,e^\prime)$ scattering data on $^3He$ at sufficiently high
momentum transfer \cite{SiDaMc-80}.

Recently new experimental data of inclusive ($e, e^\prime$) reactions
at high momentum transfer \cite{Se-al-89}, covering the $\Delta$-region
as well, have become available.
To briefly review the experimental situation, we plot in fig. 1
in the ($k$,$\omega$)-plane the
kinematics of the experiments considered in the present paper
\cite{Se-al-89,He-al-74,Ba-al-83,Me-al-84,Me-al-85}
together with the particle-hole and $\Delta$-hole response regions:
the new SLAC data clearly cover the unexplored region of
the $\Delta$-peak at high momentum transfer.

In this paper we shall show that the experimental data can be reorganized
to display a scaling property in the $\Delta$-region (we shall call
this property $y_\Delta$-scaling), once
a suitable scaling function and scaling variable are introduced.
Both experimental data and theoretical models (see Sect.~\ref{sect2})
suggest that a $y_\Delta$-scaling could occur indeed. The $y$-scaling on
the quasi-elastic peak (QEP) is usually interpreted, in fact, as an
evidence of quasi-free scattering off a nucleon inside the nucleus
when probed with a high momentum projectile: in other words, in that
kinematical situation the Impulse Approximation (IA) holds. On the other hand
one could reasonably ask if, or under which conditions,
a $\Delta$-resonance can propagate freely in the nuclear matter.
If and when such an occurrence
is met, then $y_\Delta$-scaling is expected.

We wish to give an answer to this question in the following.
But the present discussion displays from the very beginning, together
with analogies, also conceptual differences between $y$- and
$y_\Delta$-scaling. In fact the former is known to occur at high
transferred momentum because only in that kinematical
region the nucleons inside the
nuclear matter can be regarded as quasi-free. We have, instead, only
theoretical predictions about the
$\Delta$-resonance, which, {\em a priori,} could be considered
quasi-free and structureless in quite different kinematical regions.
However, the exact meaning of ``quasi-free'' and ``structureless'' has
not yet been established. Some hints on this topic will come from the
simple model of Sect.~\ref{sect2}.

We now briefly review the $y$-scaling idea, outlining analogies and
differences between the quasi-elastic and the $\Delta$-peak situations.

To begin with, $y$-scaling requires two definitions: the scaling
function and the scaling variable. To introduce the former, one needs
to know the free electron scattering
off protons ($\sigma_{ep}$) and neutrons ($\sigma_{en}$),
in terms of which the $(e,e^\prime)$ cross section reads
\cite{DaMcDoSi-90,CiPaSa-91}
\begin{equation}
\frac{d^2\sigma}{d\omega d\Omega}=\left(Z\sigma_{ep}+N\sigma_{en}\right)
K(k,\omega)F(k,\omega)\,.
\label{yd1}
\end{equation}
$K$ is a purely kinematical factor which takes
different forms in the literature
\cite{We-75,SiDaMc-80,PaSa-82,Ro-80,RiRo-87} and, to exemplify,
in the nonrelativistic limit it reads $K=k/m$; $F$ is the
scaling function. In the high momentum region $F$
becomes a function of one variable only, namely the scaling
variable. Again a variety of definitions are available.
Following for instance the Rome group\cite{CeCiSa-89}, $y$ comes out from the
energy conservation for quasielastic scattering
of an electron off a bound nucleon with minimal values for the momentum
and removal energy. The defining equation within a relativistic
kinematics reads then
\begin{equation}
\omega+M_A=\sqrt{M^2+(y+k)^2}+\sqrt{M^2_{A-1}+y^2}\;.
\label{yd2}
\end{equation}
We observe that the energy difference between the system with $A$
nucleons in the ground state and the excited system with
$A-1$ nucleons occurs, which can be interpreted as
the excitation energy of a hole in nuclear matter. In other words
the quasiparticle dynamics intervenes in defining $y$, as
we shall show in Sect.~\ref{sect2}.

Once these definitions are established, the occurrence
of $y$-scaling, i.e. the dependence of $F$ upon $y$ only, means that
only the nucleonic part of the current is effective and that the IA
mechanism works satisfactorily. Thus one is addressed to examine large
transferred $q$ and one finds that a scaling property occurs for
$y<0$, i.e. these kinematical conditions allow the
description of the nucleus as an ensemble of free quasiparticles.
In this context the analogy with the Bjorken
$x$-scaling, occurring when the nucleon constituents are seen as
structureless by the external probe, is striking. In fact the only
required input for a calculation
is now the momentum distribution of the nucleons inside
the nucleus, which is thus immediately related via IA to the
experimental data. We also remark that in the high momentum
range $y$ can be immediately
related to the Nachmann variable $\xi=(k-\omega)/M$ used in the domain
of high energy physics \cite{Na-73,GePo-76}.

There is another source of information connected with $y$-scaling,
namely the  analysis of scattering data at not too high momentum transfer:
one investigates not the occurrence, but the violation of the
$y$-scaling. This can be done, as suggested in \cite{CiPaSa-91}, by
analyzing $F$ at fixed $y$ as a function of $1/k$:
the data dispose along a straight line, whose slope is directly
connected with the nucleon-nucleon effective interaction in the nuclear
medium. Thus the information coming from the {\em violations} of the
scaling law represents an alternative point of view in using
$y$-scaling.

We now turn to the $\Delta$ peak: in this case one cannot easily factorize
the single particle process, as in (\ref{yd1}),
because a poor amount of experimental
information is available on the direct $\Delta$ electroexcitation
\cite{Ba-al-75,Ba-al-77}; moreover the presence of the nuclear medium
allows other (semi-direct) many-body processes which can lead
to the excitation to a $\Delta$ of a nucleon different from the one
struck by the incoming photon. Thus the
parametrization (\ref{yd1}) becomes meaningless on the
$\Delta$-peak. Nevertheless, we can still look for a factorization of the
cross section in the form
\begin{equation}
{d^2\sigma\over d\omega d\Omega}=A(k,\omega)K(k,\omega)F_\Delta(y_\Delta)
\label{yda1}
\end{equation}
where $A(k,\omega)$ contains information about the excitation mechanism
and $F_\Delta$ describes the ``quasi-free'' propagation of the $\Delta$
in the nuclear medium, if it can occur.

In this paper we shall derive the form of $y_\Delta$ following the same
line of thought as in the original paper of West \cite{We-75} and extending
it to the $\Delta$-peak.
The form of $A(k,\omega)$ will be fitted in order
to reach a convincing $y_\Delta$-scaling.

If we succeed in achieving this scaling behaviour, then the
information on nuclear dynamics we can extract from it are also
different. Instead of looking for the nucleon momentum distribution
-- which on the other hand is accessible through the usual $y$-scaling -- we
can investigate both the possibility of ``quasi-free'' $\Delta$ propagation
and its excitation mechanism in the nuclear medium.

In Sect.~\ref{sect2} we shall examine the elementary model of Free Fermi
Gas (FFG) to conjecture the form of the scaling variable. We shall see there
how to overcome the difficulties arising from the structure of the
$\Delta$ in the vacuum. In Sect.~\ref{sect3} we shall show how to apply
the same concept to the practical case, once suitable
modifications in the scaling function definitions are introduced.
We discuss in Sect.~\ref{sect4} the
phenomenological relevance of this kind of analysis.

\section{The Fermi Gas Model \label{sect2}}

The Free Fermi Gas model provides
a natural guide to $y$-scaling. We review these ideas, because
they suggest the extension to the $\Delta$ case.

\subsection{The $y$-scaling in the quasi-elastic peak\label{sect2.1}}

The connection between the cross section and the
response functions reads
\begin {equation}
{d^2 \sigma \over d\Omega d\epsilon'} = \sigma_{\rm Mott}
 \left\{ \left[{k^2_\mu\over k^2}\right]^2S_L (k,\omega) +
\left[\tan^2{\theta\over 2}-{k^2_\mu\over k^2}\right]S_T (k,\omega) \right\}
\label {ys-8}
\end {equation}
with $k^2_\mu=-Q^2=\omega^2-k^2$.

Consider, to face with a simple example, the longitudinal part of the
current. Assuming for the moment nonrelativistic pointlike nucleons,
the longitudinal response in the FFG model reads
\begin{equation}
S_L(k,\omega)=-{1\over \pi}\Im
{\int \frac{d^3q}{(2\pi)^3}}
{\theta(k_F-q)
{\theta\left(\left|{\bf q}+{\bf k}\right|-k_F\right)}
\over \omega-{k^2\over 2m}-{{\bf k}\cdot{\bf q}\over m}+i\eta}\;.
\label{yd3}
\end{equation}
It immediately follows that for $k>2k_F$ the function
\begin{equation}
F={k\over m}S_L(k,\omega)
\label{yd4}
\end{equation}
{\em scales}, i.e. it depends upon
\begin{equation}
y={m \omega\over k}-{k\over 2}
\label{yd5}
\end{equation}
only. If we account for the size of nucleons, then we merely factorize
the e.m. form factors.
The latter are experimentally well determined
properties of the nucleon and do not contain information about the
nuclear dynamics. Thus it is natural to redefine $F$ as
\begin{equation}
F={1\over G_E^2(Q^2)}{k\over m}S_L(k,\omega)
\label{yd6}
\end{equation}
and $F$ still scales with respect to the same variable.

In an extended FFG model we can consider the nucleus as an ensemble of
noninteracting quasi--particles. We further assume that the nucleon
self--energy in the medium is either negligible
or, at least, weakly dependent upon momentum and density.
We outline this assumption because it cannot
be automatically translated to the $\Delta$-resonance.
The nucleon self--energy then can be
approximated by means of an effective mass and a
binding effect for the nucleons below the Fermi sea. In such a case we
replace $S_L$ with
\begin{equation}
S_L(k,\omega)=-{1\over \pi}\Im{\int \frac{d^3q}{(2\pi)^3}}
{\theta(k_F-q)
{\theta\left(\left|{\bf q}+{\bf k}\right|-k_F\right)}
\over \omega-{({\bf k}+{\bf q})^2\over 2m}+{q^2\over
m^*}+\epsilon+i\eta}\;:
\label{yd7}
\end{equation}
here the scaling property is in principle destroyed, but if we consider
negligible the quantity
$$q^2/2m-q^2/2m^*$$ for $q<k_F$, an approximate scaling law still
exists, with
\begin{equation}
y={m\over k}(\omega-\epsilon)-{k\over 2}\,.
\label{yd8}
\end{equation}
Numerically this scaling property is well reproduced indeed.
{}From this schematic analysis we infer the following items:
\begin{enumerate}
\item in the longitudinal $(e,e^\prime)$ scattering the reaction
mechanism on the single nucleon is well known and factorized;
\item the factor $K(k,\omega)$ depends upon the kinematics (here
$K=k/m$);
\item the form of $y$ depends upon the dynamics of the nucleons in the
nuclear medium (effective mass and average binding).
\end{enumerate}

\subsection{The $y$-scaling in the $\Delta$-peak\label{sect2.2}}

Now we consider, at the same (low) degree of complications, the
$(e,e^\prime)$ scattering in the $\Delta$-region.
First of all we remind that the cross section is almost completely
transverse. Furthermore let us assume the
simplifying hypothesis of ``$\Delta$-dominance''
\cite{KoMo-79,OsWe-81} -- i.e. that
the whole cross section in the vacuum
can be approximated by a resonant structure. This choice is only due to
a simplicity requirement, since we shall see in Sect.~\ref{sect3} that
it is not compatible with phenomenological outcomes. It however
naturally suggests the formalism.

The previous assumption amounts to write the
whole current in the form
\begin{equation}
{\bf j}=i \tilde f_{\gamma N\Delta}{\bf S}\cdot{\bf k}\times\bfm\epsilon
\label{yd8.1}
\end{equation}
where $\tilde f_{\gamma N\Delta}$ is a suitably renormalized $\gamma
N\Delta$ coupling
constant, accounting also  for the nonresonant background.
The free $\Delta$--propagator in the vacuum is then written as
\begin{equation}
G_\Delta(k,\omega)={1\over {\omega- {{k}^2
 \over {2m_\Delta}} -
m_\Delta + m + {i \over 2} \Gamma({ k},\omega)}}
\label{yd9}
\end{equation}
where
\begin{equation}
\Gamma = \frac{2}{3}{f^2_{\pi N\Delta}\over 4\pi}{m_\Delta\over E_k m^2}
{k^*}^3 v^2_{\pi N\Delta} (k^*)
\label{yd10}
\end {equation}
$k^*$ being the relative 3-momentum of the $\pi$-nucleon outgoing pair,
defined as
\begin{equation}
k^*({\bf k},\omega)=\sqrt{m_\Delta\over E_k}
\sqrt{(E_k-m_\Delta)^2-m^2}
\label{yd10.1}
\end{equation}
with $E_k=\sqrt{(m_\Delta^2+k^2)+\omega^2}$.
$v^2_{\pi N\Delta} (k^*)$ is the form factor of the $\pi N\Delta$ vertex.
Eq. (\ref{yd10}) expresses the
dominance of the $\pi$-$N$ component in the $\Delta$-propagation, which
induces a nontrivial $k$- and $\omega$-dependence in $\Gamma$ -- the
real part of the self-energy can instead be neglected because it mainly
renormalizes the $\Delta$ mass.

Thus we see that, at variance of the QEP case,
there are no {\em a priori} reasons to neglect
the internal structure of the particle.
Nevertheless we do that for the moment and
we consider the transverse response in the nuclear medium. It reads
\begin{equation}
S_T(k,\omega)=-{1\over\pi}
{16\over 9}k^2\tilde f^2_{\gamma N\Delta}\Im
{\int \frac{d^3q}{(2\pi)^3}}
{\theta (k_F - q) \over \omega- {({\bf k + q})^2 \over
2 m_\Delta}
+ {{\bf q}^2 \over 2 m} - m_\Delta + m + i \eta}\,.
\label{yd11}
\end{equation}
Now we remark that
\begin{equation}
{{q^2} \over {2m}}-{{q^2} \over {2m_\Delta}}\ll m_\Delta - m \label{ys-5}
\end {equation}
because $q < k_F$. So again a scaling variable is naturally suggested:
we put
\begin{equation}
y_\Delta = {{m_\Delta (\omega- m_\Delta + m)} \over k} -{k \over 2} \label
{ys-6}
\end {equation}
and we conclude that
\begin{equation}
F_\Delta={1\over k}S_T=-{1\over \pi}
\tilde f^2_{\gamma N\Delta}\Im{\int \frac{d^3q}{(2\pi)^3}}
{\theta (k_F - q) \over y_\Delta-{\widehat\bfm{k}\cdot\bfm{q}\over
m_\Delta}+i\eta}
\label{yd12}
\end{equation}
is again a function of $y_\Delta$ only.
Since (\ref{ys-5}) is numerically very well verified,
(\ref{yd12}) scales even better than (\ref{yd3}) or (\ref{yd7}),
because the condition $k> 2k_F$ is not required. The presence of an
electromagnetic form factor $F_{\gamma N \Delta}(k_\mu^2)$ obviously does not
destroy the argument here presented, but simply requires to modify the
definition of the scaling function $F_\Delta$:
\begin{equation}
F_\Delta=\frac{\displaystyle 1}{\displaystyle
k F_{\gamma N \Delta}(k_\mu^2)^2}
S_T=-{1\over \pi}
\tilde f^2_{\gamma N\Delta}\Im{\int \frac{d^3q}{(2\pi)^3}}
{\theta (k_F - q) \over y_\Delta-{\widehat\bfm{k}\cdot\bfm{q}\over
m_\Delta}+i\eta}
\label{yd12a}
\end{equation}
[see, for comparison, eq. (\ref{yd6}) for the case of West scaling
and the definitions we are going to use, eqns.~(\ref{ys-9}),
(\ref{ys-9bis})].

The conclusion of this section is that a scaling variable and a scaling
function are naturally
suggested on the same grounds of the West's scaling. We are faced
however with two unjustified assumptions, namely the $\Delta$-dominance
hypothesis and the neglecting of the $\Delta$-width in the vacuum. The
former question will be discussed in the next section.
Before going on, however, we examine in some details the effect of the
$\Delta$-width.

The expression (\ref{yd12}) can be easily evaluated analytically if the
$\Delta$-width is absent\cite{BrWe-75}; to account for $\Gamma$
we invoke the low density
approximation and we replace the arguments of $\Gamma$ with the
external momentum and energy, so coming back to the previous case but with
a complex energy. Now we carry out an imaginary
experiment, assuming that the response function is correctly described
by (\ref{yd11}); we replace the experimental values obtained in
the various $(e,e')$ experiments
\cite{Se-al-89,He-al-74,Ba-al-83,Me-al-84,Me-al-85} (points of fig. 1)
with the values coming from this ``theoretical'' calculations
and we plot them as a function of $y_\Delta$ only --
with, Fig. 2a, and without, Fig. 2b, the $\Delta$-width.
Fig. 2a shows that the approximation (\ref{ys-5}) works fairly well,
while Fig. 2b proves that the free $\Delta$-width cannot be neglected
indeed and that its presence destroys the scaling property.

Thus the $y$-scaling concept seems not useful for the $\Delta$ peak
from the very
beginning, because the $\Delta$ is not structureless in the vacuum.
However many theoretical
calculations concerning the inner structure of the $\Delta$ in the
medium \cite{VaBaScWa-81,CeDi-83,OsSa-87}
suggested indeed that $\Gamma$ is strongly modified: not only its
value at the mass shell increases, but also its $k$- and
$\omega$-dependence should change -- in \cite{BrWe-75} is argued that in the
medium $\Gamma\sim {k^*}^2$ instead of ${k^*}^3$.
Would the $\Delta$-width become roughly
constant in the medium near the mass-shell, then a scaling property could
still occur.

\section{Evidence of $y_\Delta$ scaling\label{sect3}}

In the previous section we saw how the FFG model suggested, under two severe
hypotheses, a scaling behaviour for experimental data in the $\Delta$--region.
Now we ask if and how these indications are realized by phenomenology.

We shall not deal further, in this paper, with other possible choices for
$y_\Delta$ -- as a minimal requirement, we should account for
relativistic kinematics -- delaying these problems to future investigations;
instead, we look for an appropriate form of
$F_\Delta$. We shall proceed by induction, starting from the up to now
unjustified $\Delta$-dominance hypothesis, and adding those corrections
that are suggested by phenomenology.

To construct the scaling function we remind, first of all, that
in the $\Delta$-region $S_L$   is likely negligible,
so that the kinematical factor in front of $S_T$ can be factorized, and
the same happens for the $\gamma N\Delta$ form factor. Furthermore,
a factor $k^2$ coming from the e.m. vertex (\ref{yd8.1})
must also be accounted for. In passing, we
note that in a truly relativistic calculation of the vertex
part of the polarization propagator $k^2$ is replaced by  $-k^2_\mu$
\cite{AlCe-93}. We shall adopt this indication in the remaining of this
paper.

On these grounds the following scaling function is suggested:
\begin{equation}
F_\Delta(y_\Delta,k)
={1 \over A}\;{k \over (-k^2_\mu)
{\sigma_M(k_\mu^2)F_{\gamma N\Delta} (k^2_\mu)}}\;
{1 \over {\tan^2{\theta\over 2}-{k^2_\mu\over k^2}}}
\;{{d^2 \sigma} \over {d \Omega d \epsilon'}} \label {ys-9}\;.
\end {equation}
It is worth comparing this expression with (\ref{yd1}). So far a strict
analogy survives: in fact
$$(-k^2_\mu)F_{\gamma N\Delta}(k_\mu^2)
\left[{\tan^2{\theta\over 2}-{k^2_\mu\over
k^2}}\right]\sigma_M(k_\mu^2)$$
describes the direct $\Delta$-electroexcitation
on a nucleon in the vacuum and $k$ is the analogous of the factor
$K(k,\omega)$ in (\ref{yd1}). Since, moreover,
it is known that the $\Delta$ electroproduction scales with $A$
\cite{Oc-al-84,CeChDi-85}, we introduced in (\ref{ys-9}) the factor $1/A$
in order to compare different nuclei.

Eq. (\ref{ys-9}) however introduces a relevant source of uncertainties,
because the $N$-$\Delta$ transition form factor is largely unknown.
In fact typical expressions coming from the constituent quark model
are able to reproduce the ratio between the e.m. nucleon form factor and
the $N$-$\Delta$ transition one, but they usually provide a too strong cut
at high $k$ for both of them\cite{Gi-91}; on the other hand, few
$N(e,e^\prime\pi)N^\prime$ experimental data are
available\cite{Ba-al-75,Ba-al-77}.
Thus we have chosen for convenience the form factor
\begin{equation}
F_{\gamma N\Delta} (k^2_\mu) =
\left(\frac{\Lambda_1^2-(M_\Delta-M)^2}{\Lambda_1^2-k_\mu^2}\right)^2
\left(\frac{\Lambda_2^2}{\Lambda_2^2-k_\mu^2}\right)^2
\end{equation}
with $\Lambda_1 = 710. Mev/c$, $\Lambda_2 = 1200 MeV/c$, following the
naive idea that the e.m. size of the $\Delta$ should be somehow larger than
the one of the nucleon. Other choices, with different normalizations, do
not significantly alter the present considerations.

Next we consider the set of data of refs.
\cite{Se-al-89,He-al-74,Ba-al-83,Me-al-84,Me-al-85}.
For each point in the $(k,\omega)$ plane we evaluated the corresponding
$y_\Delta$ according to (\ref{ys-6}) and we plotted (\ref{ys-9}) as a
function of $y_\Delta$ only.
The results are presented in Fig.~3. We remember that with our definitions
the $\Delta$-peak is more or less centered around $y_\Delta=0$, the region
of the resonance being $-0.2\stackrel{\le}{\sim}y_\Delta
\stackrel{\le}{\sim} 0.2$, see, e.g., Fig.~2a: it is evident that no
convincing scaling behaviour arises. It is remarkable in Fig.~3 the
unfamiliar shape that the $\Delta$-peak takes: in fact each set of
experimental points disposes
itself along lines which do not present anymore a peak; this because
we do not plot directly the cross sections, but the
scaling function (\ref{ys-9}). The $k_\mu^2$ in the denominator of
eq.~(\ref{ys-9}) decreases as long as $k_0$ increases, so enhancing the
values of the scaling function for larger $y_\Delta$. These considerations
hold true for Figs.~5,6 and 7 as well, where a more convincing scaling
behaviour arises.

We could of course ascribe the absence of scaling in fig.~3
to the $\Delta$-width structure,
but before giving up $y_\Delta$-scaling we explore if other reaction
mechanisms can lead to a resonant behaviour. This analysis
is suggested by a careful comparison of Figs.~2b and 3: the presence of
a finite and $k$ dependent width of the resonance actually
destroys the scaling property of the free Fermi gas scaling function,
but the  order of magnitude of the expected violations are completely different
- of the order of 2 at $y_\Delta=0$ in Fig.~2b against a factor of the
order of 10 or more at $y_\Delta=0$ in Fig.~3.
As a matter of fact, two mechanisms at least
are known to be relevant in the $\pi$-photoproduction reaction,
namely the direct $\Delta$ photoexcitation and the Kroll-Rudermann term
\cite{BlLa-77}, the second being dominant at the threshold for charged pions
production. So far we accounted, implicitly, only for the first one. The
second one -- it can occur in the forward scattering amplitude only due
to the presence of the nuclear medium -- creates a pion plus a particle-hole
pair (see fig. 4b): the pion brings with itself almost all the
transferred momenta,  as it is suggested by the static limit for
nucleons,
and the p-h pair brings with itself the helicity of the transverse photon.
Of course, once the pion is created inside nuclear matter, it can
resonate with a $\Delta$-hole pair, thus providing another resonant
mechanism, competitive with the direct $\Delta$-excitation, which is
forbidden in the vacuum. We clearly do not know a priori, without
modelizing, if such a mechanism can be relevant, even if we obviously
expect it to be more and more important when moving to the left of
the $\Delta$-peak (in the low energy region). An insight about this,
however, comes from the ratio between $\tilde f_{\gamma N \Delta}$, the
$\Delta$-dominance model effective coupling constant, and
$f_{\gamma N \Delta}$, the c.c. as it comes out from pion photoproduction
experiments. According to \cite{BlLa-77,OsWe-81} this ratio is $\sim 1.5$.

As already emphasized, our aim is not to provide a microscopic
calculation, but instead to reorganize experimental data
to be in closer contact with theoretical previsions. So, if our
conjecture is correct -- i.e. if the direct electro-excitation plus the
Kroll-Rudermann term truly dominate the cross section -- it is
natural to replace the factor $k_\mu^2$ in the denominator of eq.~(\ref{ys-9})
with
\begin{equation}
k^2_\mu\longrightarrow
\left\{\alpha k^2_\mu+{\beta k_\mu^2\over
{\left(k^2_\mu-m_\pi^2\right)^2}}\right\}
\end{equation}
(again a relativistically correct calculation of
the $\Delta$-h excitation induced by a pion requires to replace the
vertex ${k}^2$ with $-k_\mu^2$\cite{CeSa-91}), where the coefficients
$\alpha$ and $\beta$ are related to the relative weights of these
terms. Thus we replace (\ref{ys-9}) with
\begin{equation}
F_\Delta={1 \over A}{k \over \sigma_M F_{\gamma N\Delta} (k^2_\mu)}
{1\over \alpha k^2_\mu+{\beta k_\mu^2\over \left(k^2_\mu-m_\pi^2\right)^2}}
{1 \over
\tan^2{\theta\over 2}-{k^2_\mu\over k^2}}
{d^2 \sigma \over d \Omega d \epsilon'} \label {ys-9bis}
\end {equation}
and we fit the parameters $\alpha$ and $\beta$ -- of course only
the ratio between them being meaningful. The results are plotted in fig.~5,
with $m_N^4\alpha/\beta=17.95$. We see now a good scaling property
in the region $-0.2<y_\Delta<0.2$, just the whole region of the
$\Delta$-peak in fig. 2a. Two important remarks are here necessary.
\begin{enumerate}
\item As widely discussed in the Introduction a conceptual difference
arises between $y$ and $y_\Delta$--scaling: in fact $y$ scaling occurs
in the region of large negative $y$ as well, outside the QEP, and
provides, there, information on the nucleon momentum distributions.
The $y_\Delta$-scaling, instead, is seen to occur just in the region of the
peak (the whole region however), while the inner structure of
the $\Delta$, resumed in the complicated momentum and energy dependence
of its width, is presumably disturbing outside this region.
This difference reflects itself into the different
kinds of information we can extract from the two scaling properties.
\item The direct electro-excitation seems to be dominant.
The Kroll-Ru\-der\-mann term is however important near the pion mass
shell, so mainly affecting the low momentum region of the $\Delta$
peak.
\end{enumerate}

A further improvement can be achieved if we allow further rescattering
of the pion in nuclear matter. Considering, for sake of simplicity, two
rescattering at most, in fig. 6  we plotted the data divided by
\begin{equation}
\alpha k^2_\mu+{\beta k_\mu^2\over \left(k^2_\mu-m_\pi^2\right)^2}
+{\gamma k_\mu^4\over \left(k^2_\mu-m_\pi^2\right)^3}\;\;.
\label{ys9ter}
\end{equation}
This time we find $m_N^4\alpha/\beta=5.9$, but
$\beta\sim-\gamma$.

As a by-product,
we remark also that in fig. 6 experiments on different nuclei are reported.
This by one side proves again that these data scale with $A$ too,
as supposed since long time and as sum rules suggest \cite{CeChDi-85};
on the other hand this procedure can weaken the scaling property with
$y_\Delta$, because different experiments can have different systematical
errors.

The plot of fig.~6 is our central result. To understand its meaning we
compare it with the results of our ideal experiment of fig.~2b. In fig.~6 we
fitted the three parameters $\alpha$, $\beta$ and $\gamma$ to reduce as far
as possible the spreading of the experimental data at $y_\Delta=0$. Even so
the points are still somehow spread out, mainly because different
experiments have been collected together. In fact fig~7,
where only the SLAC data on $^{12}C$ are reported, shows a better
scaling behaviour indeed. Of course further refinements can be achieved by
improving the definition of the scaling variable and function.

It is impressive in fig.~6 the fact the scaling is realized within the same
accuracy on the whole range of the resonance, namely $0.2<y_\Delta<0.2$.
This outcome must be compared with the one of fig.~2b: there data, in
absence of experimental noises,
concentrated in one point at $y_\Delta\simeq -0.23$, but for higher values
of $y_\Delta$ their spreading is sensibly higher with respect to fig.~6.

A further comment concerns the differences between figs.~5 and 6: fig.~5
displays a few points scattered out of the central line, a feature
disappeared in fig.~6. In practice, at a first sight, only a small
improvement is achieved, but when looking to the values of the fitted
parameters we realize that they significantly change and the outcome
$\beta\sim-\gamma$ strongly support the idea of a resonant pion
propagation.

The last open question concerns the effect of the $\Delta$-width.
The occurrence of a scaling effect seems to suggest that $\Gamma$
really changes when plunged in the nuclear medium. How can we
corroborate this insight?

A comparison between experimental data on a nucleus and on a free proton
provides the desired indications.
To minimize noises and uncontrolled systematical errors, we selected the data
taken from only one laboratory \cite{Se-al-89} and only one nucleus
(${}^{12}C$) and we plotted
them together with the $p (e e^\prime) X$ data from the same laboratory
-- having assumed $\beta=\gamma=0$ for the free scattering, since in the vacuum
the corresponding processes are not allowed.
The results are given in fig. 7 and look very impressive:
{\em the scaling property is fulfilled on a nucleus, but not on
the single nucleon}. This last negative result is obviously expected due to
the non trivial energy and momentum dependence of the $\Delta$ width in the
vacuum. But the order of magnitude of the scaling violation in this case
compared with the $^{12}C$ result gives us an insight on the influence of
the medium in flattening the momentum and energy dependence of the $\Delta$
width.

Thus $y_\Delta$-scaling analysis suggests, from experimental data, that
the drastic $k$- and $\omega$-dependence of the $\Delta$-width is
indeed smoothed by the nuclear medium.

\section{Discussions and Perspectives\label{sect4}}

We derived, in this paper, the gross structures of $y_\Delta$
scaling, which can be summarized as follows:
\begin{enumerate}
\item A scaling property in the $\Delta$ peak arises if
the experimental data are collected in terms of $y_\Delta$ eq. (\ref{ys-6}).
\item The scaling property is strongly sensitive to the excitation mechanisms
of the resonance.
\end{enumerate}
Of course many topics need further careful investigations. Let us
outline the main open problems.

\paragraph{The definition of $y_\Delta$.}

We derived $y_\Delta$ from the FFG model. Two
improvements are required: by one side we should consider the possible
effect of an average binding felt by the $\Delta$ in the nuclear medium,
but this is probably compensated by other effects (short range
correlations and $\rho$-meson rescattering over the $\Delta$-hole pair)
so that the position of the peak remains more or less unchanged
\cite{CeDi-84}, thus we do not expect relevant corrections from this
effect. On the other side relativistic corrections to the form
of $y_\Delta$ are surely relevant, in particular for the data at high
momentum transfer, like those from SLAC.

\paragraph{The Kroll-Rudermann term.}

The need for this contribution is clear from the comparison
of fig. 3 with figs. 5 and 6. However, by one side microscopic
calculations of the excitation mechanism
are wanted, on the other our approach can be
further improved by replacing the vacuum pion propagators in (\ref{ys-9bis}),
(\ref{ys9ter}) with the correspondig ones
in the medium. Reorganizing low and medium
momentum transfer data according
to $y_\Delta$-scaling can help in better understanding the relative effect
of direct electroexcitation and contact term and also in enlightening
the open problem of the pion propagation in the medium.

\paragraph{The $\Delta$-width in the medium.}

We discussed before the effect of the free $\Delta$-width, and we
concluded
that it leads to a poor scaling. Actually the scaling seems to be
satisfactorily verified. However, should this achievement be ascribed to
the fitting procedure? The answer is, in our opinion, negative. In fact
the effect of the $\Delta$ width is relevant at high momentum transfer,
while in the same region the contact terms are more or less negligible.
Thus another indication can be extracted from experimental data, namely
that the energy and momentum dependence of $\Gamma$ is flattened in the
medium, due to the opening of many--body channels forbidden in the
vacuum: this reflects into a milder threshold effect.

\paragraph{The electromagnetic form factor.}

We have outlined that the $\gamma N \Delta$ transition form factor is
poorly known and only few attempts are available in giving realistic
theoretical descriptions \cite{WaScPfRo-90}: we resorted to use, at
least, a reasonable expression. However, once the $y_\Delta$ scaling is
seen to occur, and remarking again that the Kroll-Rudermann terms are
negligible at high momentum, the possibility is offered for extracting
the value of the form factor from experimental data.

The Kroll-Rudermann term should also be multiplied by a form factor, but
this is a priori different from that of the direct electro-excitation
term: it instead can be derived from the $\pi NN$ form factor by means
of the Ward-Takahashi identities.

Needless to say, each of these points requires further, careful
investigations.

\newpage
\centerline{{\bf\Large Figure captions}}

\begin{enumerate}
\item Plot in the $(k,\omega)$ plane of the experimental kinematics
considered in the present work. Boxes correspond to the data of ref.
\cite{Se-al-89}, diamonds to ref. \cite{He-al-74},
circles to ref. \cite{Ba-al-83} and dots to refs. \cite{Me-al-84,Me-al-85}.

\item FFG scaling functions without (a) and with (b) the $\Delta$ width.

\item $F_\Delta$  within the $\Delta$-dominance hypothesis
(eq. (\ref{ys-9})) as a function of $y_\Delta$ alone  for all nuclei.

\item The direct (a) and Kroll-Rudermann (b) mechanisms able to excite a
$\Delta$--resonance in nuclear matter, together with the corresponding
many--body diagrams.

\item $F_\Delta$ with direct plus Kroll-Rudermann term
(eq. (\ref{ys-9bis})) as a function of $y_\Delta$ alone  for all nuclei

\item $F_\Delta$ with direct plus Kroll-Rudermann term and rescattering
(eq. (\ref{ys9ter})) as a function of $y_\Delta$ alone  for all nuclei

\item $F_\Delta$ as a function of $y_\Delta$ alone. SLAC
experiment only, $p(e,e^\prime)X$ and $^{12}C$.
\end{enumerate}

\begin{thebibliography}{10}

\bibitem{DaMcDoSi-90}
{D. B. Day, J. S. McCarthy, T. W. Donnelly and I. Sick}.
\newblock {\it Ann. Rev. Part. Sci.}, 40:357, 1990.

\bibitem{CiPaSa-86}
{C. Ciofi degli Atti, E. Pace and G. Salm\`e}.
\newblock {\it Few Body Syst.}, Suppl. 1:280, 1986.

\bibitem{CiPaSa-91}
{C. Ciofi degli Atti, E. Pace and G. Salm\`e}.
\newblock {\it Phys. Rev.}, C43:1155, 1991.

\bibitem{We-75}
G.~B. West.
\newblock {\it Phys. Rep.}, C18:263, 1975.

\bibitem{SiDaMc-80}
{I. Sick, D. Day and S. J. McCarthy}.
\newblock {\it Phys. Rev. Lett.}, 45:871, 1980.

\bibitem{Se-al-89}
R.~M.~Sealock et~al.
\newblock {\it Phys. Rev. Lett.}, 62:1350, 1989.

\bibitem{He-al-74}
F.~H.~Heimlich et~al.
\newblock {\it Nucl. Phys.}, A231:509, 1974.

\bibitem{Ba-al-83}
P.~Barreau et~al.
\newblock {\it Nucl. Phys.}, A402:515, 1983.

\bibitem{Me-al-84}
Z.~E.~Meziani et~al.
\newblock {\it Phys. Rev. Lett.}, 52:2180, 1984.

\bibitem{Me-al-85}
Z.~E.~Meziani et~al.
\newblock {\it Phys. Rev. Lett.}, 54:1233, 1985.

\bibitem{PaSa-82}
{E. Pace and G. Salm\`e}.
\newblock {\it Phys. Lett.}, 110B:411, 1982.

\bibitem{Ro-80}
{R. Rosenfelder}.
\newblock {\it Ann. Phys.}, 128:188, 1980.

\bibitem{RiRo-87}
{A. S. Rinat and R. Rosenfelder}.
\newblock {\it Phys. Lett.}, B193:411, 1987.

\bibitem{CeCiSa-89}
{R. Cenni, C. Ciofi degli Atti and G. Salm\`e}.
\newblock {\it Phys. Rev.}, C39:1425, 1989.

\bibitem{Na-73}
{O. Nachman}.
\newblock {\it Nucl. Phys.}, B63:237, 1973.

\bibitem{GePo-76}
{H. Georgi and H. D. Politzer}.
\newblock {\it Phys. Rev.}, D14:1829, 1976.

\bibitem{Ba-al-75}
{G. Bardin et al.}
\newblock {\it Nuovo Cimento Lett.}, 13:485, 1975.

\bibitem{Ba-al-77}
{G. Bardin et al.}
\newblock {\it Nucl. Phys.}, B120:45, 1977.

\bibitem{KoMo-79}
{I. H. Koch and E. J. Moniz}.
\newblock {\it Phys. Rev.}, C20:235, 1979.

\bibitem{OsWe-81}
{E. Oset and W. Weise}.
\newblock {\it Nucl. Phys.}, A368:375, 1981.

\bibitem{BrWe-75}
{G. E. Brown and W. Weise}.
\newblock {\it Phys. Rep.}, C22:281, 1975.

\bibitem{VaBaScWa-81}
{J. W. Van Orden, M. K. Banerjee, D. M. Schneider and S. J. Wallace}.
\newblock {\it Phys. Rev.}, C23:2157, 1981.

\bibitem{CeDi-83}
{R. Cenni and G. Dillon}.
\newblock {\it Nucl. Phys.}, A392:438, 1983.

\bibitem{OsSa-87}
{E. Oset and L. L. Salcedo}.
\newblock {\it Nucl. Phys.}, A468:631, 1987.

\bibitem{AlCe-93}
{W. M. Alberico and R. Cenni}.
\newblock 1993.
\newblock Work in progress.

\bibitem{Oc-al-84}
{J. S. O'Connel et al.}
\newblock {\it Phys. Rev. Lett.}, 53:1627, 84.

\bibitem{CeChDi-85}
{R. Cenni, P. Christillin and G. Dillon}.
\newblock {\it Phys. Lett.}, B151:5, 1985.

\bibitem{Gi-91}
{M. M. Giannini}.
\newblock {\it Rep. Progr. Phys.}, 54:453, 1991.

\bibitem{BlLa-77}
{I. Blomqvist and J. M. Laget}.
\newblock {\it Nucl. Phys.}, A208:405, 1977.


\bibitem{CeSa-91}
{R. Cenni and P. Saracco}.
\newblock {\it Nuvo Cimento}, A104:821, 1991.

\bibitem{CeDi-84}
{R. Cenni and G. Dillon}.
\newblock {\it Nucl. Phys.}, A422:527, 1984.

\bibitem{WaScPfRo-90}
{M. Warns, H. Schr\"oder, W. Pfeil and H. Rollnik}.
\newblock {\it Z. Phys.}, C45:627, 1980.

\end{thebibliography}
\end{document}